\documentclass[prd,superscriptaddress,floatfix,nofootinbib,twocolumn,aps]{revtex4-1}
\usepackage{graphicx,amssymb,amsmath,bm,dcolumn,xcolor}
\usepackage{threeparttable}

\usepackage[bottom]{footmisc}
\usepackage{comment}
\usepackage{xfrac}
\usepackage{appendix}
\usepackage{enumitem}
\usepackage[normalem]{ulem}

\definecolor{dukeblue}{rgb}{0.0, 0.0, 0.65}

\begin{document}
\title{Searching for scalar dark matter with compact mechanical resonators }

\author{Jack Manley}
\affiliation{Department of Electrical and Computer Engineering, University of Delaware, Newark, DE 19716, USA}

\author{Dalziel J. Wilson}
\affiliation{College of Optical Sciences, University of Arizona, Tucson, AZ 85721, USA}

\author{Russell Stump}
\affiliation{Department of Electrical and Computer Engineering, University of Delaware, Newark, DE 19716, USA}

\author{Daniel Grin}
\affiliation{Department of Physics and Astronomy, Haverford College, Haverford, PA 19041, USA}

\author{Swati Singh}
\affiliation{Department of Electrical and Computer Engineering, University of Delaware, Newark, DE 19716, USA}
\email{swatis@udel.edu}

\begin{abstract}

Ultralight scalars are an interesting dark matter candidate which may produce a mechanical signal by modulating the Bohr radius. Recently it has been proposed to search for this signal using resonant-mass antennae. Here, we extend that approach to a new class of existing and near term compact (gram to kilogram mass) acoustic resonators composed of superfluid helium or single crystal materials, producing displacements that are accessible with opto- or electromechanical readout techniques. We find that a large unprobed parameter space can be accessed using ultra-high-Q, cryogenically-cooled, cm-scale mechanical resonators operating at 100 Hz to 100 MHz frequencies, corresponding to $10^{-12}-10^{-6}$ eV scalar mass range.
 
\end{abstract}

\maketitle

%\section{Introduction}
%-----------------------------------------------------------
\textit{Introduction}.--The existence of dark matter (DM) is supported by numerous astrophysical observations~\cite{1970ApJ...159..379R,Tyson:1998vp, Markevitch:2003at,Hinshaw:2012aka,Aghanim:2018eyx}. However, the Standard Model (SM) of particle physics provides no clear DM candidates, spurring searches for new (beyond the SM) particles like WIMPs (weakly interacting massive particles) \cite{Jungman:1995df,Tan:2016zwf, Akerib:2016vxi} and axions \cite{Peccei:1977hh,Wilczek:1977pj,Weinberg:1977ma,Kim2010}. String theory suggests many new light particles, motivating the possibility of ultralight dark matter \cite{Witten:1984dg,Damour:1994ya,Damour:1994zq,Svrcek:2006yi,Conlon:2006tq,Arvanitaki:2009fg}.

For sufficiently low masses ($m_{\text{dm}}\lesssim 10^{-1}~{\rm eV}$),  DM particles behave as a classical field, due to their large occupation numbers. DM would then be produced non-thermally through coherent oscillations of a cosmological scalar field \cite{Abbott:1982af,Dine:1982ah,PhysRevD.28.1243,Preskill:1982cy}. Cosmic microwave background anisotropies, large-scale structure observations, and other measurements impose a lower limit of $m_{\text{dm}}\gtrsim 10^{-22}~{\rm eV}$ for ultralight DM (c.f. \cite{Hlozek:2014lca,Marsh:2015xka,Hlozek:2017zzf,Poulin:2018dzj,Irsic:2017yje,Kobayashi:2017jcf,Armengaud:2017nkf,Gonzales-Morales:2016mkl}).

Under a parity transform, some ultralight DM particles (such as axions) transform as pseudoscalars, while others (e.g. dilatons and moduli) transform as scalars. The parameter space for new ultralight scalars has been constrained by stellar cooling bounds~\cite{Hardy:2016kme,Graham:2015ouw} and by torsion balance experiments \cite{adelberger2009torsion,Wagner:2012ui}. Through couplings to the SM, scalar fields would modulate the fine-structure constant $\alpha$ and lepton masses (e.g. the electron mass $m_{e}$). \cite{Damour:2002mi,Damour:2010rp}. If this scalar field is the dark matter, this modulation would occur at the DM Compton frequency, $\omega_{\text{dm}}=m_{\text{dm}}c^{2}/\hbar$, an effect
 detectable using atomic clocks, atom interferometry, laser interferometry,  and other methods~\cite{Arvanitaki:2014faa,Stadnik:2014tta,Stadnik:2015kia,Stadnik:2015uka,Stadnik:2016zkf,Arvanitaki:2015iga,Arvanitaki:2017nhi}. 
 
 Modulation of $\alpha$ and $m_{\rm e}$ also produces a mechanical signal---an oscillating atomic strain---through modulation of the Bohr radius, $a_0=\hbar/\alpha c m_{\rm e}$ \cite{Arvanitaki:2015iga}. This strain can give rise to measurable displacement in a body composed of many atoms, and be resonantly enhanced in an elastic body with acoustic modes at $\omega_\text{dm}$. Recently it has been suggested to search for this \textit{acoustic} DM signature using resonant-mass antennae  \cite{Arvanitaki:2015iga}. Data from the AURIGA gravitational wave (GW) detector has already put bounds on scalar DM coupling  \cite{Branca:2016rez}. In Ref.~\cite{Arvanitaki:2015iga}, new resonant DM detectors were proposed, including a frequency-tunable Cu-Si sphere coupled to a Fabry-P\'{e}rot cavity, and more compact quartz bulk acoustic wave (BAW) resonators  \cite{2013NatSR...3E2132G}. A technique for broadband detection of low mass scalar DM was explored in Ref.~\cite{Geraci:2018fax}.

Here we propose extending the compact-resonator approach to a broader class of existing gram to kilogram-scale devices composed of superfluid He or single crystals. These devices (along with BAW resonators discussed earlier~\cite{Arvanitaki:2015iga}) have been studied in the field of cavity optomechanics \cite{DeLorenzo2017,Rowan2000,Neuhaus2017cooling}, and provide access to a broad frequency (mass) range from $100~{\rm Hz}\lesssim \omega_{\text{dm}}/2\pi\lesssim 100~{\rm MHz}$ ($10^{-12}~{\rm eV}\lesssim m_{\text{dm}}\lesssim 10^{-6}~{\rm eV}$). The key virtue of this approach is that, owing to their small dimensions and crystalline material, these devices can be operated at dilution refrigerator temperatures with quality factors as high as $10^{10}$ \cite{2013NatSR...3E2132G}, thereby substantially reducing thermal noise. We present analytic expressions for thermal-noise-limited DM sensitivity for an arbitrary acoustic mode shape, and find that the minimum detectable scalar coupling can be orders of magnitude below current bounds.

 \begin{figure*}[ht]
	\begin{center}
		\includegraphics[width=2.0\columnwidth]{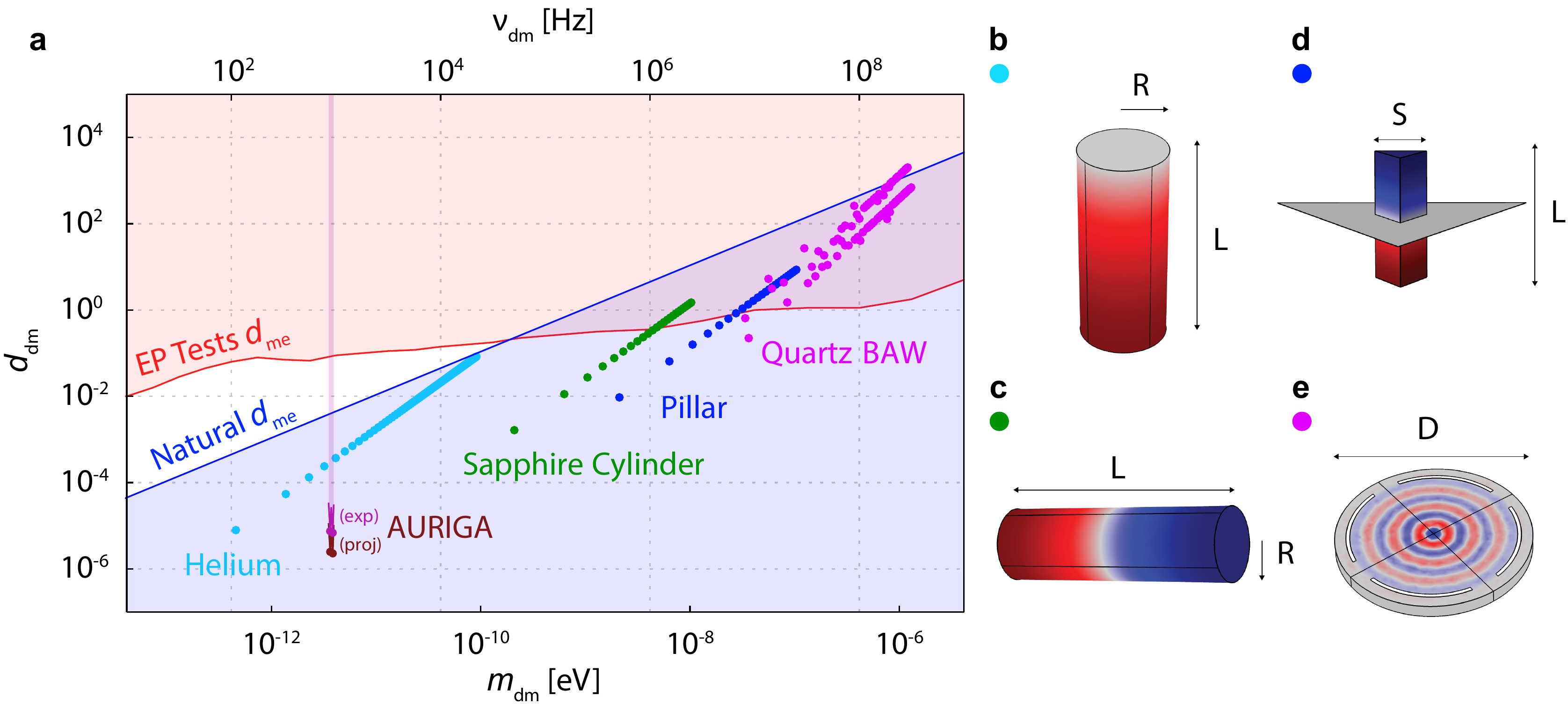}
		\caption{{\bf{(a)}} Log-log plot of coupling strength $d_{\text{dm}}$ versus DM frequency $\nu_{\text{dm}}$ and mass $m_{\text{dm}}$, assuming $d_e=0$. The red region is excluded by the E\"{o}t-Wash EP test\cite{Wagner:2012ui,Arvanitaki:2015iga}. Further constraint is provided by analysis of $\sim 1$ month of data by AURIGA (exp). The projected sensitivity of AURIGA for a full $\sim \!10$ years of signal integration is in burgundy\cite{Branca:2016rez}. The blue region is natural for electron Yukawa coupling with a 10 TeV cut-off \cite{Arvanitaki:2015iga}. Solid circles are the predicted minimum detectable coupling $\left(d_{\text{dm}}\right)_{\text{min}}$ for each proposed detector, assuming an integration time of 1 year and experimental parameters described in the main text. Light blue points: $\left(d_{\text{dm}}\right)_{\text{min}}$ for the first 100 longitudinal modes of a superfluid helium detector.  Green points: $\left(d_{\text{dm}}\right)_{\text{min}}$ for the first 25 odd-ordered longitudinal modes of a cylindrical HEM$^\circledR$ sapphire test mass \cite{Rowan2000}. Dark blue points: $\left(d_{\text{dm}}\right)_{\text{min}}$ for the first 25 odd-ordered longitudinal modes of a sapphire micropillar \cite{Neuhaus2017}. Lavender points: $\left(d_{\text{dm}}\right)_{\text{min}}$ for lower order longitudinal modes of quartz BAW resonators \cite{Goryachev:2014yra,2013NatSR...3E2132G}. {\bf{(b)}} Rendering of superfluid helium detector. Following the design in Ref. \cite{Singh:2016xwa}, we use: $R=10.8$ cm, $L=50$ cm. {\bf{(c)}} Rendering of HEM$^\circledR$ sapphire test mass. From Ref.~\cite{Rowan2000} $R=15$ mm, $L=10$ cm. {\bf{(d)}} Rendering of sapphire pillar. $s=4$ mm, $L=1$ cm. {\bf{(e)}} Rendering of quartz BAW resonator. From Ref.~\cite{Goryachev:2014yra,2013NatSR...3E2132G} Device 1: $L_1=1$ mm, $D_1=30$ mm, $R_{1}=300$ mm. Device 2: $L_2=1.08$ mm, $D_2=13$ mm, $R_{2}=230$ mm. $L$ is the thickness and $R$ is the radius of curvature of the top surface. 
		}
		\label{fig:plot}
	\end{center}
	\vspace{-3mm}
\end{figure*}

%\section{Scalar DM coupling}	\label{DMproperties}
%-------------------------------------------------------------------
\textit{Scalar DM field properties}--DM particles in the Milky Way have a Maxwellian velocity distribution about the virial velocity $v_{\text{vir}}\approx10^{-3}  c$~\cite{Derevianko:2016vpm}. Given the local DM density ($\rho_{\text{dm}}\approx 0.3$ GeV/cm\textsuperscript{3}~\cite{Lewin:1995rx}), ultralight DM particles behave as a classical field. We consider DM as a field with coherence time $\tau_{\text{c}}=\left(\frac{{v_{\text{vir}}}^2}{c^2} \omega_{\text{dm}}\right)^{-1}$ and coherence length $\lambda_\text{c}$ equal to the de Broglie wavelength $\lambda_\text{dm}$ \cite{Derevianko:2016vpm}. DM mass $m_{\text{dm}}\lesssim 10^{-6}~{\rm eV}$ corresponds to $\lambda_{\text{dm}}\gtrsim 1$ km, implying that the field is spatially uniform over laboratory scales.

Coupling of dark matter to $\alpha$ and $m_e$ leads to an oscillating strain given by~\cite{Arvanitaki:2015iga}
\begin{equation}
h(t)=-\frac{\delta \alpha\left(t\right)}{\alpha_0}-\frac{\delta m_e\left(t\right)}{m_{e,0}}=-h_0 \cos{\left(\omega_\text{dm}t\right)},
\end{equation}
where 
\begin{equation}
    h_0=d_\text{dm} \sqrt{\frac{8\pi G \rho_\text{dm}}{{\omega_\text{dm}}^2 c^2}}.
\end{equation}
Here $d_{\text{dm}}=d_{m_e}+d_e$ is a dimensionless constant describing the strength of the DM coupling to the electron mass ($d_{m_e}$) and fine-structure constant ($d_e$)~\cite{Damour:2010rp,Arvanitaki:2014faa, Arvanitaki:2015iga}.

%\section{Resonant mass detection}	
%-----------------------------------------------------------
\textit{Resonant mass detection}.--A scalar DM field modulates the size of atoms (by $h$, fractionally) at the Compton frequency $\omega_{\text{dm}}$. This effect introduces an isotropic stress in a solid body (rather, any form of condensed phase matter). This stress is effectively spatially uniform over length scales much smaller than $\lambda_\text{c}$  \cite{Derevianko:2016vpm}. Such a periodic stress may excite acoustic vibrations in the body. Note that not every acoustic mode couples to DM; a point that we wish to emphasize is that a uniform  stress only couples to \textit{breathing modes}.

Mechanical resonators that operate in non-breathing modes are not sensitive to scalar DM strain. An example of modes that would not be excited are those of a rigidly clamped solid bar. In this case, a spatially uniform stress will not cause any of the atoms in the bar to displace from their equilibrium position because of the zero net force on each. Without rigid clamping to impose an equal and opposite force on the edges of the bar, the bar will be free to expand and contract. We have found that by introducing at least one free acoustic boundary, a spatially uniform stress can couple to acoustic modes. It is for this reason that we specify that only breathing modes couple to scalar DM.  
 
To quantify the effect of DM on an elastic body (the detector), we have adapted the analysis for continuous gravitational waves in Ref.~\cite{Hirakawa1973}. We begin with the displacement field $u_i=\sum_n{\xi_n(t)u_{ni}(\boldsymbol{x})}$, where $u_{ni}$ is the normalized spatial distribution and $\xi_n$ is the time-dependent amplitude of the $n$\textsuperscript{th} acoustic mode; subscript $i$ denotes the spatial component \{$x$,$y$,$z$\}. This allows us to model the detector as a harmonic oscillator with effective mass $\mu_n=\int\rho\sum_i\left|u_{ni}\right|^2dV$. It is driven by thermal forces, $f_{\text{th}}(t)$, and a DM-induced force, $f_{\text{dm}}(t)=\ddot{h}(t)q_n$, where $q_n=\int\rho\sum_iu_{ni}x_idV$ is a parameter that determines the strength of the coupling between a scalar strain and the $n^\text{th}$ mode of the detector. By introducing dissipation in the form of velocity damping, the modes of the resonator obey damped harmonic motion
\begin{equation}	\label{EoM}
\ddot{\xi}_n+\frac{\omega_n}{Q_n}\dot{\xi}_n+\omega_n^2\xi_n
=\frac{f_{\text{dm}}}{\mu_n}+\frac{f_{\text{th}}}{\mu_n},
\end{equation}
where $\omega_n$ and $Q_n$ are, respectively, the resonance frequency and quality factor of the $n$\textsuperscript{th} mode.

Thus, the strategies developed for resonant detection of gravitational waves, originally proposed by Weber \cite{Weber1960}, can also be applied to detecting DM~\cite{Arvanitaki:2015iga}. Note that not all GW detectors double as scalar DM detectors. Broadband interferometric detectors, such as LIGO, are only sensitive to gradients in the DM strain field \cite{Arvanitaki:2014faa}. A spatially uniform isotropic strain would produce equal phase shifts in each arm of an interferometer. Moreover, scalar DM strains atoms, not free space---in this sense it is not equivalent to a scalar GW.

\begin{figure}[t]
	\begin{center}
		\includegraphics[width=1.0\columnwidth]{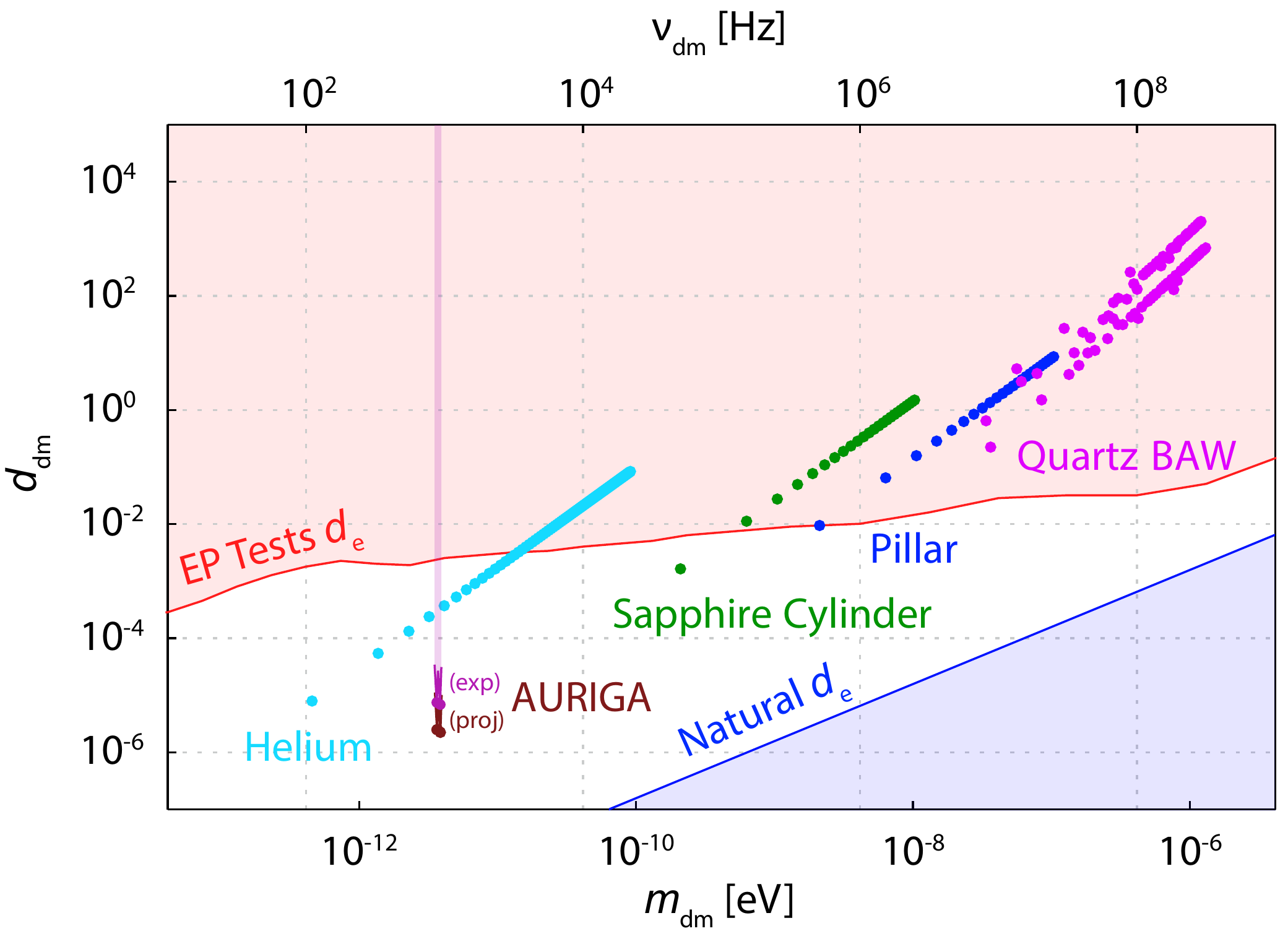}
		\caption{Coupling strength $d_{\text{dm}}$ vs DM frequency $\nu_{\text{dm}}$ and mass $m_{\text{dm}}$ in $d_e$ parameter space. Point types and colors are as in Fig. 1. Higher sensitivities are needed to probe new parameter space for $d_e$ coupling than for $d_{m_e}$. 
		}
		\label{fig:plotde}
	\end{center}
	\vspace{-3mm}
\end{figure}

%\section{DM Parameter Space}	
%-----------------------------------------------------------
\textit{DM Parameter Space}.--The parameter space for scalar couplings $d_{m_e}$ and $d_{e}$ is shown in Figs. \ref{fig:plot} and \ref{fig:plotde}, respectively. Each plot includes sensitivity estimates for four candidate detectors (discussed below and in the caption). Overlaid are experimental constraints set by EP tests (the E\"{o}t-Wash experiment) and gravitational wave searches (AURIGA), as well as the benchmark ``natural $d_{\rm dm}$" line. Below we briefly review these constraints.

The E\"{o}t-Wash experiment, a long-standing test of the weak equivalence principle using a torsion balance, has set the strongest existing constraints on $d_{m_e}$ and $d_e$. The orange exclusion region in Fig. \ref{fig:plot}(a) comes from the comparison of the differential accelerations of beryllium and titanium masses to $10^{-13}$ precision \cite{Wagner:2012ui}. 

AURIGA is a resonant-mass gravitational wave detector based on a $3$-m-long, $2200$ kg Al-alloy (Al5056) bar cooled to liquid He temperatures \cite{Branca:2016rez}. The detector has collected $\sim \! 10$ years of data, one month of which has been analyzed to search for scalar DM \cite{Branca:2016rez}. Extrapolating to its full (10 year) run time, the DM sensitivity of AURIGA is $\left(d_{\text{dm}}\right)_{\text{min}}\approx 10^{-5}$ for $850\text{Hz}\leq \nu_{\text{dm}} \leq 950 \text{Hz}$. This bandwidth is set by the sensitivity over which thermal motion of the Al bar can be detected.

 The naturalness criterion requires that quantum corrections to $m_{\text{dm}}$ be smaller than $m_{\text{dm}}$ itself \cite{Dimopoulos:1996kp}. Consistent with other work~\cite{Dimopoulos:1996kp, Arvanitaki:2016fyj,Arvanitaki:2015iga}, this cutoff is chosen as roughly the energy scale up to which the SM is believed to be valid. The blue region in Fig. \ref{fig:plot} indicates where the naturalness criterion is satisfied for a cutoff of 10 TeV.

%\section{Thermal noise and minimum detectable coupling}	
%-----------------------------------------------------------
\textit{Thermal noise and minimum detectable coupling}.--Mechanical strain sensors, like AURIGA, are fundamentally limited by thermal noise.  We consider mm to cm-scale mechanical resonators operating at Hz to MHz frequencies, for which thermal motion is the dominant noise source but deep cryogenics and quantum-limited displacement readout are available. The expression for thermally-limited strain sensitivity was first applied to resonant-mass DM detection in Ref. \cite{Arvanitaki:2015iga}. Here, we summarize the derivation of strain sensitivity, arriving at general expressions for arbitrary resonator geometries.

Thermal noise is well-described by a white-noise force spectrum, $S_{ff}^{\text{th}}=\frac{4k_{\text{B}}T\mu_n\omega_n}{Q_n}$, which drives the mechanical resonator into Brownian motion \cite{Saulson:1990jc}. Following Eq. (\ref{EoM}), this limits the sensitivity of a strain measurement to
\begin{equation}    \label{strainsensitivity}
\sqrt{S_{hh}^{\rm th}}=\sqrt{\frac{4k_{\text{B}}T\mu_n}{Q_n{q_n}^2{\omega_n}^3}}.
\end{equation}

Accounting for the DM field's finite coherence time, the minimum detectable strain for $2\sigma$ detection of the signal over measurement duration $\tau_{\text{int}}\gg\tau_\text{c}$ is
\begin{equation} \label{hmin}
h_{\text{min}}\approx\sqrt{\frac{16{v_{\text{vir}}} k_{\text{B}}T\mu_n}{Q_n{q_n}^2{\omega_n}^{5/2}c}}{\tau_{\text{int}}}^{-\frac{1}{4}}.
\end{equation}
The minimum detectable DM coupling is
\begin{equation}	\label{dmin}
\left(d_{\text{dm}}\right)_{\text{min}}\approx \sqrt{\frac{2{v_{\text{vir}}}  c}{\pi G \rho_{\text{dm}}}}\sqrt{\frac{k_{\text{B}}T\mu_n}{Q_n{q_n}^2\sqrt{\omega_n \tau_{\text{int}}}}},
\end{equation}
which can also be expressed in terms of the minimum detectable strain as
\begin{equation} \label{dminhmin}
\left(d_{\text{dm}}\right)_{\text{min}}\approx\sqrt{\frac{c^2}{8\pi G \rho_{\text{dm}}}}\omega_n h_{\text{min}}.
\end{equation}

Equations (\ref{strainsensitivity})-(\ref{dminhmin}) are analytical expressions, general to any mechanical detector of arbitrary elastic material and geometry. Equation (\ref{dmin}) is used to generate the results for each detector in Fig. \ref{fig:plot}(a) for $\tau_{\text{int}}=1$ year.

Typical $h_{\text{min}}$ values derived for the devices in this work are $\sim10^{-24}-10^{-23}$. From Eq. (\ref{dminhmin}) it is evident that higher frequency detectors require a lower $h_{\text{min}}$ in order to maintain the same minimum detectable coupling. This scaling arises from the inverse relationship between the DM field amplitude $h_0$ and Compton frequency $\omega_\text{dm}$. 

 Another challenge to high frequency detection is that the DM signal's coherence time $\tau_\text{c}$ is inversely proportional to the Compton frequency. Rearranging Eq. \eqref{hmin} gives (for $\tau_\text{int}\gg\tau_\text{c}$) $h_{\text{min}}=2\sqrt{S_{hh}^{\rm th}}\left(\tau_{\text{int}}\tau_{\text{c}}\right)^{-1/4}$. Thus, a shorter coherence time increases $(d_\text{dm})_\text{min}$. 
 
 The detector geometry also introduces unfavorable frequency scaling, as higher frequency resonators are generally smaller, implying a reduced coupling factor $q_n$. Geometric considerations reduce $q_n$ for higher $n$ modes.
 
 For the reasons explained above, $\left(d_{\text{dm}}\right)_{\text{min}}$ tends to scale as $\sim\omega_{\text{dm}}^{7/4}$ for simple, longitudinal modes. Thus, designing mechanical resonators to beat limits set by EP tests is difficult in the $\omega_\text{dm}\sim\,\text{GHz}$ range.

%\section{Device Parameters and Results} 
%-------------------------------------------------------------------
\textit{Device parameters and results}.--We now consider several possible scalar dark matter detectors based on acoustic breathing mode resonators. Figure \ref{fig:plot} highlights four resonators with gram to kilogram effective masses and Hz-MHz frequencies. Each detector behaves like a miniature Weber Bar antenna \cite{Branca:2016rez}. To facilitate comparison, we assume a 10 mK operating temperature and mechanical Q-factors of $10^9$, unless otherwise constrained by experiment. Specific parameters are stated in the caption of Fig. \ref{fig:plot}. Note that while the mode shapes in Fig. \ref{fig:plot}(b-e) are rendered numerically in COMSOL$^\circledR$ \cite{ComsolMultiphysics}, the results plotted in Fig. \ref{fig:plot}(a) and Fig. \ref{fig:plotde} are analytical. 

For DM frequencies $100 \, \text{Hz}\lesssim \nu_{\text{dm}} \lesssim 25 \, \text{kHz}$, we consider the superfluid helium bar resonator probed optomechanically, as discussed in Ref.~\cite{DeLorenzo2017} (Fig. \ref{fig:plot}(b)). To permit breathing modes, the helium container designed to be only partially filled. The niobium shell supporting the container is assumed to be infinitely rigid due to its much greater bulk modulus. The resonant medium is the $2.7$ kg volume of superfluid. Assuming $T=10$ mK and $Q=10^9$ (limited by doping and clamping loss)  \cite{DeLorenzo2017}, $\left(d_{\text{dm}}\right)_{\text{min}}$ for the first 100 longitudinal modes is plotted in light blue in Fig. \ref{fig:plot}(a). For the fundamental mode ($\nu_1\approx120$ Hz), the strain sensitivity is $\sqrt{S_{hh}^{\rm th}}=2.5\cdot 10^{-21}$ Hz\textsuperscript{-1/2}.

For DM frequencies $50\,\text{kHz}\lesssim \nu_{\text{dm}} \lesssim 2.5 \, \text{MHz}$, we consider a $0.3$ kg HEM$^\circledR$ sapphire cylinder intended for use as an end-mirror in future cryogenic GW detectors~\cite{Rowan2000}. We note that an existing class of similar, promising devices are not considered in this work \cite{Locke1998,Locke2000,nand2013resonator,Hirose:2014xga,Bourhill2015}. We assume $T=10$ K as an experimental constraint due to the low thermal conductance of the test mass suspensions \cite{Khalaidovski:2014fqa}.  A quality factor of $Q=10^9$ is assumed based on historical measurements of Braginsky \emph{et. al.} \cite{Bagdasarov1975,Braginsky1985systems}, though we note a more contemporary benchmark is $Q = 2.5\times 10^8$ at $T = 4 $ K \cite{Uchiyama:1999ne}. Green points in Fig. \ref{fig:plot}(a) are estimates of $\left(d_{\text{dm}}\right)_{\text{min}}$ for $25$ longitudinal modes with dimensions as shown in Fig. \ref{fig:plot}(c). For the fundamental mode ($\nu_1\approx54$ kHz) the strain sensitivity is $\sqrt{S_{hh}^{\rm th}}=2.4\cdot 10^{-22}$ Hz\textsuperscript{-1/2}.

For DM frequencies $550\,\text{kHz}\lesssim \nu_{\text{dm}} \lesssim 27 \, \text{MHz}$, we consider a modification of the quartz micropillar resonator developed by Neuhaus \emph{et. al.} \cite{Neuhaus2017,Neuhaus2017cooling} (see also Ref. \cite{Kuhn2011micropillar})  for cryogenic optomechanics experiments.  The micropillar is assumed to be scaled up in size (Fig. \ref{fig:plot}(d)) and reconstructed of sapphire, whose higher density and sound velocity produces larger strain coupling in order to begin ruling out parameter space in the MHz regime with only $\sim 0.3$ grams of mass. Estimates of $\left(d_{\text{dm}}\right)_{\text{min}}$ for the first 25 odd-ordered longitudinal modes, with $Q=10^9$ and $T=10$ mK, are shown in blue in Fig. \ref{fig:plot}(a). For the fundamental mode ($\nu_1=550$ kHz), the strain sensitivity is $\sqrt{S_{hh}^{\rm th}}=7.7\cdot 10^{-23}$ Hz\textsuperscript{-1/2}.

Finally, for DM frequencies $10\,\text{MHz}\lesssim \nu_{\text{dm}} \lesssim 350 \, \text{MHz}$, we consider two gram-scale quartz BAW resonators \cite{2013NatSR...3E2132G}, initially proposed to search for scalar DM in Ref.~\cite{Arvanitaki:2015iga}. Lavender points in Fig. \ref{fig:plot}(a) are for several longitudinal modes assuming an average quality factor of $10^{10}$ for Device 1 and $10^9$ for Device 2, with $Q$ adjusted for a few specific modes corresponding to measurements in Ref. \cite{2013NatSR...3E2132G}. Due to the unfavorable frequency scaling described above, these BAWs are predicted to surpass $d_{m_e}$ EP test constraints for only a few lower order modes, when operating at $T=10$ mK. The strain sensitivity for the mode at $\nu\approx10$ MHz is $\sqrt{S_{hh}^{\rm th}}\approx5\cdot 10^{-23}$ Hz\textsuperscript{-1/2} . 

Excluded from the figures are high frequency devices such as phononic crystals \cite{chan2012optimized,maccabe2019phononic} and GHz BAWs \cite{renninger2018bulk}.  We found them unable to compete with EP test constraints. In principle one could extend our work to lower frequency mechanical resonators. In this case sensitivity would ultimately be limited by strain noise due to Newtonian gravity gradients and seismic fluctuations \cite{Adhikari:2013kya}. 

\textit{Detector readout requirements and bandwidth}.--We have considered the thermal limit to resonant-mass DM detection for various compact resonators.  To reach this limit, the imprecision of the readout system $S_{hh}^\text{imp}$ must be smaller than thermal noise $S_{hh}^\text{th}$, yielding a fractional detection bandwidth of $\Delta \omega/\omega\approx Q^{-1}\sqrt{S_{hh}^\text{th}/S_{hh}^\text{imp}}$. 

The resonators discussed permit high-sensitivity optomechanical readout. Sapphire cylinders and pillars can be mirror-coated (e.g. using crystalline coatings \cite{cole2013tenfold}) and coupled to a Fabry-P\'{e}rot cavity. For devices in Fig. 1, thermal displacement of the end-face is on the order of $10^{-14}\,\text{m}/\sqrt{\text{Hz}}$ (cylinder) and  $10^{-16}\,\text{m}/\sqrt{\text{Hz}}$ (pillar) near the fundamental resonance, implying a fractional bandwidth of $10^{-5}$ ($10^{-7}$) for a shot-noise-limited displacement sensitivity of $10^{-18}\,\text{m}/\sqrt{\text{Hz}}$ (achievable with mW of optical power for a cavity finesse of $1000$).

Superfluid-He and quartz BAW resonators have been probed non-invasively with low-noise microwave circuits. The piezoelectricity of quartz permits contact-free capacitive coupling of a BAW to a superconducting quantum interference device (SQUID) amplifier; this has enabled fractional bandwidths of $10^{-6}$ for a 10 mK, 10 MHz with $Q\sim 10^8$ device \cite{goryachev2014observation}. Helium bars have likewise been capacitively coupled to superconducting microwave cavities. For the bar considered in Fig. \ref{fig:plot}, a detailed roadmap to thermal-noise-limited readout is described in Ref. \cite{Singh:2016xwa}.

Frequency tuning can also increase the effective detector bandwidth. The sound speed of quartz and sapphire are both thermally tunable, however, ultra-cryogenic operation practically limits the utility of this approach. Superfluid He permits broadband mechanical tuning by pressurization (which has been used to change the sound speed of He by 50\% \cite{Abraham:1969zz}).  Another possible route is through dynamical coupling to the microwave or optical resonator used for readout.  Though weak, such ``optical spring" effects (well studied in cavity optomechanics \cite{Aspelmeyer:2013lha}) are noninvasive and might be used to trim the detector at the level of the fractional DM signal bandwidth, $\Delta \omega_\text{dm}/\omega_\text{dm}=(\omega_\text{dm}\tau_\text{c})^{-1}\sim 10^{-6}$.

Tradeoffs between bandwidth, sensitivity and tunability ultimately determine the search strategy for a given detector. For instance, while three of the detectors discussed above (based on helium bar, sapphire cylinder and sapphire micropillar resonators) can surpass the sensitivity of the E\"{o}t-Wash experiment in under a minute, their bandwidth will likely be smaller than that of the DM signal $\Delta \omega_\text{dm}$. To widen the search space, a natural strategy (analogous to haloscope searches for axion DM) would be to scan the detector in steps of $\Delta\omega_\text{dm}$, each time integrating for a duration long enough to resolve thermal noise $\tau_\text{int} \gtrsim 4Q/\omega_\text{dm}\times S_{hh}^\text{imp}/S_{hh}^\text{th}$. The slow scaling of sensitivity with $\tau_\text{int}$ (Eq. \ref{hmin}) allows this strategy to significantly enhance the effective detector bandwidth. The total run time of the experiment can be reduced (or bandwidth increased) by using more detectors, which is facilitated by the compactness of the devices proposed.

\textit{Conclusion and outlook}.--Existing, or near term compact mechanical resonators with high quality-factor acoustic modes operating at cryogenic temperatures have the potential to beat constraints on DM-SM coupling strength set by tests for EP violations in the 100 Hz- 100 MHz range. Frequency tuning techniques, along with arrays of these compact resonators can be used to enhance bandwidth and sensitivity, thereby enabling table-top experiments to cover a vast, unexplored region in the DM-SM coupling parameter space.

%\begin{acknowledgments}
We thank Keith Schwab, David Moore, Andrew Geraci, Michael Tobar, and Eric Adelberger for helpful conversations. We thank Ken Van Tilburg, Asimina Arvanitaki, and Savas Dimopoulos for extensive feedback on the manuscript, as well as stimulating conversations. This work is supported by the National Science Foundation grant PHY-1912480, and the Provost's Office at Haverford College.
%\end{acknowledgments}

\appendix
\section{Scalar DM coupling}	\label{DMproperties}

Here we review how scalar DM would interact with Standard Model fields through terms in which gauge-invariant operators of a SM field are coupled to operators containing DM fields\textcolor{red}~ {\cite{Arvanitaki:2014faa,Derevianko:2016vpm}, following the notation of Ref.~\cite{Derevianko:2016vpm}}.

We begin by considering only linear couplings, denoted by Lagrangian density $\mathcal{L}_{\rm lin}=\sqrt{\hbar c}\phi({\bf x},t) \sum_x \gamma_x \mathcal{O}_{\rm SM}$,
where $\gamma_x$ is the coupling coefficient and $\mathcal{O}_{\rm SM}$ are terms from the SM Lagrangian density. For simplicity, we consider only coupling to the electron (denoted by fermionic field $\psi_e$) and electromagnetic field strength (denoted by Faraday tensor $F_{\mu\nu}$).  Thus
\begin{equation}
-\mathcal{L}_{\rm lin}=\sqrt{\hbar c}\phi({\bf x},t) \left[-\frac{\gamma_e}{4}F_{\mu\nu} F^{\mu\nu} + \gamma_{m_e}\bar{\psi}_e\psi_e\right].
\end{equation}
Combining it with the SM Lagrangian, this coupling can be absorbed into variations of fundamental constants~\cite{Damour:2010rp}
\begin{eqnarray}
m_e ({\bf x},t)&=&m_{e,0}\left[1+\sqrt{\hbar c}\gamma_{m_e}\phi({\bf x},t)\right],\\
\alpha ({\bf x},t)&=&\alpha_{0}\left[1+\sqrt{\hbar c}\gamma_{e}\phi({\bf x},t)\right],
\end{eqnarray}

One can introduce dimensionless couplings $d_{m_e}$ and $d_e$ and consider the fractional change of constants
\begin{eqnarray}\label{eq:DMoscillations}
\frac{\delta m_e ({\bf x},t)}{m_{e,0}}&=&d_{m_e}\sqrt{4\pi \hbar c} E_{\rm Pl}^{-1}\phi({\bf x},t),\\\label{eq:DMoscillations2}
\frac{\delta \alpha ({\bf x},t)}{\alpha_{0}}&=&d_{e}\sqrt{4\pi \hbar c} E_{\rm Pl}^{-1}\phi({\bf x},t),
\end{eqnarray}
where $E_{\rm Pl}$ is the Planck energy ($E_{\rm Pl}=\sqrt{\hbar c^5/G}$) \cite{Geraci:2018fax}. 

The couplings $d_{m_e}, d_e$ are  dimensionless dilaton-coupling coefficients \cite{Damour:2010rp,Arvanitaki:2014faa, Arvanitaki:2016fyj}, with a natural parameter range defined by the inequality \cite{Arvanitaki:2015iga}
\begin{equation}
m_{\rm dm}^{2}\geq \frac{1}{2\left(4\pi\right)^{4}}d_{m_{e}}y_{e}^{2}\frac{\Lambda^{4}}{M_{\rm pl}^{2}}
+\frac{1}{32\pi^{2}}d_{e}^{2}\frac{\Lambda^{4}}{M_{\rm pl}^{2}},
\label{eq:nat}
\end{equation}where $M_{\rm pl}$ is the reduced Planck mass and $y_{e}=2.94\times 10^{-6}$ is the electron Yukawa coupling. Eq.~(\ref{eq:nat}) imposes the requirement that quantum corrections to the scalar mass be well-controlled, assuming a $\Lambda=10~{\rm TeV}$ cutoff.

\section{Minimum detectable strain and integration time} \label{timescaling}

Over a finite measurement time $\tau$, the power spectral density $S_{hh}\left(\omega\right)$ of a coherent signal  $h(t)=h_0e^{-i\omega_n t}$ has an apparent magnitude 
\begin{equation} \label{coherentsignalint}
S_{hh}^\tau(\omega_n)=\frac{1}{\tau}\left<\left|H^{\tau}(\omega_n)\right|^2\right>={h_0}^2\tau.
\end{equation}

If $h$ is partially coherent with coherence time $\tau_{\text{c}}$, then (\ref{coherentsignalint}) is only a valid approximation for $\tau<\tau_{\text{c}}$. For measurement times $\tau\gg\tau_{\text{c}}$, a better approximation can be obtained by breaking the measurement into $N$ segments of duration $\tau_{\text{c}}$ and adding up the contributions in quadrature \cite{Budker:2013hfa}.  For a stationary process, this yields
\begin{equation}
S_{hh}^{\tau\gg\tau_\text{c}}\approx\sqrt{\sum\limits^{N} \left(S_{hh}^{\tau_\text{c}}\right)^2}=\sqrt{\frac{\tau}{\tau_{\text{c}}}\left(S_{hh}^{\tau_\text{c}}\right)^2}= h_0^2\sqrt{\tau\tau_\text{c}},
\end{equation}
from which a signal strength
\begin{equation}
h_0=\sqrt{S_{hh}^\tau}\left(\tau\tau_{\text{c}}\right)^{-1/4}
\end{equation}
can be inferred.

We define the minimum detectable strain $h_{\text{min}}$ as the minimum signal amplitude $h_0$ needed to produce $\text{SNR}=1$. For $2\sigma$ detection limited by thermal noise $S_{hh}^{\text{th}}$,
\begin{equation}
h_{\text{min}}\approx2\sqrt{S_{hh}^{\text{th}}}\left(\tau\tau_{\text{c}}\right)^{-1/4}.
\end{equation}

\section{Effect of readout noise}

The preceding analysis assumes that noise in the readout (of amplitude coordinate $\xi$) contributes negligibly to the apparent strain.  In practice broadband readout noise $S_{\xi\xi}^\text{imp}(\omega)\approx S_{\xi\xi}^\text{imp}(\omega_n)$ contributes an apparent strain
\begin{equation}
S_{hh}^\text{imp}(\omega) = |\chi(\omega)|^{-2}S_{\xi\xi}^\text{imp}(\omega_n)
\end{equation}
where
\begin{equation}
|\chi(\omega)|^{2} = \frac{\omega^4 q_n^2/\mu_n^2}{(\omega^2-\omega_n^2)^2+\omega_n^2\omega^2/Q_n^2}
\end{equation}
is the mechanical susceptibility.

The effect of readout noise on a measurement of finite duration $\tau$ is obtained by integrating the readout signal $S_{\xi\xi}(\omega_n)$ over a bandwidth $\Delta \omega = 2\pi/\tau$.  For times $\tau\gg\tau_c$, the contribution of thermal and readout noise is 
\begin{subequations}
	\begin{align}
	S_{\xi\xi}^\tau(\omega_n) &= \int^{\omega_n+\tfrac{\Delta \omega}{2}}_{\omega_n-\tfrac{\Delta \omega}{2}}(S_{\xi\xi}^\text{imp}(\omega)+|\chi|^2 S_{hh}^\text{th}(\omega))\frac{d\omega}{\Delta\omega}\\
	&\approx S_{\xi\xi}^\text{imp}(\omega_n)+S_{\xi\xi}^\text{th}(\omega_n)\frac{\tan^{-1}\left(\tau_n/\tau\right)}{\tau_n/\tau}\label{eq:SNRvstime}
	\end{align}
\end{subequations}
where $\tau_n\equiv2\pi Q_n/\omega_n$ is the mechanical coherence time.

According to Eq. \ref{eq:SNRvstime}, the relative fraction of readout noise is minimized for integration times long compared to the mechanical coherence time $\tau\gg\tau_n$.  For integration times  $\tau\ll\tau_n$, relevant for frequency scanning, the fraction is $S_{\xi\xi}^\text{imp}/S_{\xi\xi}^\text{th}\times 2\tau_n/(\pi\tau)$.  We use this formula in the main text to define the time necessary to resolve thermal noise as $2\tau_n/\pi\times S_{\xi\xi}^\text{imp}/S_{\xi\xi}^\text{th} = 4Q/\omega_n\times S_{hh}^\text{imp}/S_{hh}^\text{th}$.

As a specific example, a superfluid helium resonator with the dimensions discussed in main text, probed with a signal-to-noise ratio of $\sqrt{S_{hh}^\text{th}/S_{hh}^\text{imp}}=10$ for an integration time of $\tau_\text{int}\approx15$ hours, could in two years search a fractional frequency span of $\Delta\omega/\omega_\text{dm}\approx 0.1\%$ ($\sim 10^3$ distinct bins) with a sensitivity of $(d_\text{dm})_\text{min}\sim10^{-5}$, exceeding the current bound set by EP tests by more than 20 dB.

\section{Equation of Motion}	\label{EoMderivation}

Dark matter modulates the size of atoms by $h\equiv h(t)$. In a linearly elastic medium, this effect is analogous to modulating the equilibrium position of each atom relative the center of the medium (or an edge, if that edge is clamped in place). In an isotropic medium, the effect can be modeled as a perturbation, $-x_ih$, to the displacement field, $u_i\equiv u_i({\boldsymbol{x}},t)$. The treatment follows that of Ref.~\cite{Hirakawa1973} for continuous gravitational waves.

The $i^\text{th}$ component of the perturbed displacement field is simply
\begin{equation}
w_i\equiv w_i({\boldsymbol{x}},t)=u_i({\boldsymbol{x}},t)-x_ih(t).
\end{equation}

It should here be noted that this model only strictly applies for elastic media with at least one free acoustic boundary. A bar, for example, that is rigidly clamped at one end needs to have zero displacement $w_i$=0 at the rigid boundary. The model still applies to this case, but only if the rigid boundary is positioned at the origin $x_i=0$. 

Navier's equations of motion~\cite{ContinuumMechanics} for the perturbed displacement field become
\begin{equation} \label{navier}
\rho \ddot{u}_i-\mu\sum_j\frac{\partial^2 u_i}{\partial {x_j}^2} - (\lambda + \mu) \sum_j\frac{\partial^2u_j}{\partial x_i \partial x_j}=\rho\ddot{h} x_i,
\end{equation}
where $\rho$ is the mass density of the detecting medium and $\mu$ and $\lambda$ are Lam\'{e} parameters. 

The displacement field due to acoustic oscillations can be expanded in terms of its eigenmodes: $u_i({\boldsymbol{x}},t)=\sum_n \xi_n \! (t) \, u_{ni}({\boldsymbol{x}})$, where $\xi_n\equiv \xi_n \! (t)$ gives the amplitude and phase of the oscillation while $u_{ni}\equiv u_{ni}({\boldsymbol{x}})$ is the normalized spatial distribution. The normalization is such that $(u_{ni})_{\text{max}}=1$. Without loss of generality, we can restrict our analysis to just one of the eigenmodes
\begin{equation}
u_i=\xi_nu_{ni}.
\end{equation}

With this substitution into (\ref{navier}), we recover the equation of motion for a driven, harmonic oscillator 
\begin{equation}
\mu_n\left(\ddot{\xi}_n+\omega_{n}^{2}\xi_n\right)=\ddot{h}q_n,
\end{equation}\\
where $\mu_n=\int\mathrm{d}V\rho\sum_i\left|u_{ni}\right|^2$ is the effective mass of the $n$\textsuperscript{th} mode and $q_n=\int\mathrm{d}V\rho\sum_iu_{ni}x_i$ characterizes coupling between scalar DM strain and the $n$\textsuperscript{th} mode. Not every mode will couple. We have found that only breathing modes couple to an isotropic, spatially uniform strain. 

Finally, we include velocity-proportional damping $\frac{\omega_n}{Q_n}$, and random thermal noise, $f_{\text{th}}$, and the equation of motion for the $n$\textsuperscript{th} eigenmode of the medium is
\begin{equation}
\ddot{\xi}_n+\frac{\omega_n}{Q_n}\dot{\xi}_n+\omega_n^2\xi_n
=\frac{q_n}{\mu_n}\ddot{h}+\frac{f_{\text{th}}}{\mu_n}.
\end{equation}

\section{Acoustic Analysis of Devices}	\label{ModeDetails}

Here we consider the geometries of the proposed detectors, showing the analytical values of the effective mass $\mu_n$ and acoustic coupling factor $q_n$. 

The sapphire test mass and pillar (Fig.\ref{fig:plot}(c-d)) are simple bars with free acoustic boundaries. Consider such a bar with length $L$ and cross-sectional area $A$. It's ends are located at $z=0$ and $z=L$. The longitudinal displacement modes are~\cite{kinsler1999fundamentals}
\begin{equation}	\label{longmodes}
u_{nx}=u_{ny}=0; \,\,\,\,\,\,\, u_{nz}=\cos{\!\left[\frac{n\pi z}{L}\right]}.
\end{equation}
Thus, for a bar with arbitrary cross-sectional geometry, the reduced mass is
\begin{equation}
\mu_n=\rho A \int_0^L\mathrm{d}z \cos^2{\!\left[\frac{n\pi z}{L}\right]}=\frac{M}{2},
\end{equation}
where $M$ is the total mass, and the acoustic coupling factor is
\begin{equation} \label{couplingfactorbar}
q_n=\rho A \int_0^L\mathrm{d}z \cos{\!\left[\frac{n\pi z}{L}\right]} z=\rho A L^2 \frac{\cos{\!\left(n\pi\right)}-1}{n^2 \pi^2}.
\end{equation}
Equation (\ref{couplingfactorbar}) illustrates that only the odd-ordered longitudinal modes couple to dark matter. Even-ordered modes are not breathing modes. In terms of the speed of sound in the material $v_s$, the resonance frequencies are $\nu_n=\frac{n v_s}{2L}$.

The geometry of the proposed superfluid helium cylinder in Fig.\ref{fig:plot}(b) differs only in that it has a rigid acoustic boundary at $z=0$. For this geometry, the longitudinal displacement modes are 
\begin{equation}
u_{nx}=u_{ny}=0; \,\,\,\,\,\,\, u_{nz}=\sin{\!\left[\frac{\left(2n-1\right)\pi z}{2L}\right]}.
\end{equation}
The effective mass is still
\begin{equation}
\mu_n=\frac{M}{2},
\end{equation}
and the acoustic coupling factor is now
\begin{equation}
q_n=-4\rho A L^2 \frac{\cos{\!\left(n\pi\right)}}{\left(2n-1\right)^2 \pi^2}.
\end{equation}
Modes of both even and odd $n$ couple to DM strain, and the frequency is $\nu_n=\frac{\left(2n-1\right) v_s}{4L}$.

To approximate the displacement field for the quartz BAW resonators, we assume the crystal to be only weakly anisotropic and consider only the dominant component $u_{nz}$ of the quasi-longitudinal modes. The displacement modes are given by 
\begin{equation} \label{bawdisplacement}
u_{nz}\approx\sin{\!\left[\frac{n\pi z}{L}\right]}\exp{\left[\frac{-\alpha n \pi }{2}\left(x^2+y^2\right)\right]},
\end{equation}
with frequency
\begin{equation}
\nu_n\approx\sqrt{\frac{n^2 \hat{c}_z}{4 L^2 \rho}},
\end{equation}
where $\alpha\approx\sqrt{\frac{2}{5RL^3}}$ and $\hat{c}_z$ is the effective elastic constant~\cite{Goryachev:2014yra}. From Eq.~ (\ref{bawdisplacement}), we calculate $\mu_n$ and $q_n$ for odd-ordered modes, finding that
\begin{equation}
\mu_n\approx\frac{\rho L}{2\alpha n}
\end{equation}
and
\begin{equation}
\left|q_n\right|\approx\frac{4\rho L^2}{n^3 \pi^2 \alpha}.
\end{equation}

\bibliography{DMsensor}

\end{document}